\documentclass[manuscript]{acmart}

\setcopyright{rightsretained}
\acmJournal{PACMHCI}
\acmYear{2023} \acmVolume{7} \acmNumber{CSCW1} \acmArticle{28} \acmMonth{4} \acmPrice{}

\setcopyright{none}
\renewcommand\footnotetextcopyrightpermission[1]{}
\makeatletter
\gdef\@copyrightpermission{
  \begin{minipage}{0.2\columnwidth}
   \href{https://creativecommons.org/licenses/by/4.0/}{\includegraphics[width=0.90\textwidth]{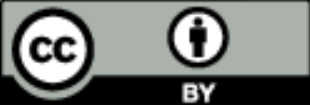}}
  \end{minipage}\hfill
  \begin{minipage}{0.8\columnwidth}
   \href{https://creativecommons.org/licenses/by/4.0/}{This work is licensed under a Creative Commons Attribution International 4.0 License.}
  \end{minipage}
  \vspace{5pt}
}
\makeatother

\usepackage[utf8]{inputenc}
\usepackage{wspr}
\usepackage{url}
\usepackage{booktabs}
\usepackage{vars}
\usepackage{amsmath}

\begin{document}

\title[Beyond the Boolean]{Beyond the Boolean: How Programmers Ask About, Use, and Discuss Gender}

\author{Elijah Bouma-Sims}
\email{eboumasi@andrew.cmu.edu}
\orcid{0000-0002-9999-6949}
\affiliation{
  \institution{Carnegie Mellon University}
  \city{Pittsburgh}
  \state{Pennsylvania}
  \country{United States of America}
}

\author{Yasemin Acar}
\email{acar@gwu.edu}
\orcid{0000-0001-7167-7383}
\affiliation{
  \institution{The George Washington University}
  \city{Washington, D.C.}
  \country{United States of America}
  }


\begin{abstract}
Categorization via gender is omnipresent throughout society, and thus also computing; gender identity is often requested of users before they use software or web services. Despite this fact, no research has explored how software developers approach requesting gender disclosure from users. To understand how developers think about gender in software, we present an interview study with \numtotalinterviews~software developers recruited from the freelancing platform Upwork as well as Twitter. We also collected and categorized \numtotalposts~threads that contained keywords relevant to gender from programming-related sub-forums on the social media service Reddit. \numrelevantposts~posts that discussed approaches to gender disclosure were further analyzed. We found that while some developers have an understanding of inclusive gender options, programmers rarely consider when gender data is necessary or the way in which they request gender disclosure from users. Our findings have implications for programmers, software engineering educators, and the broader community concerned with inclusivity.
\end{abstract}
\maketitle

\section{Introduction}

\begin{figure}
    \centering
    \includegraphics{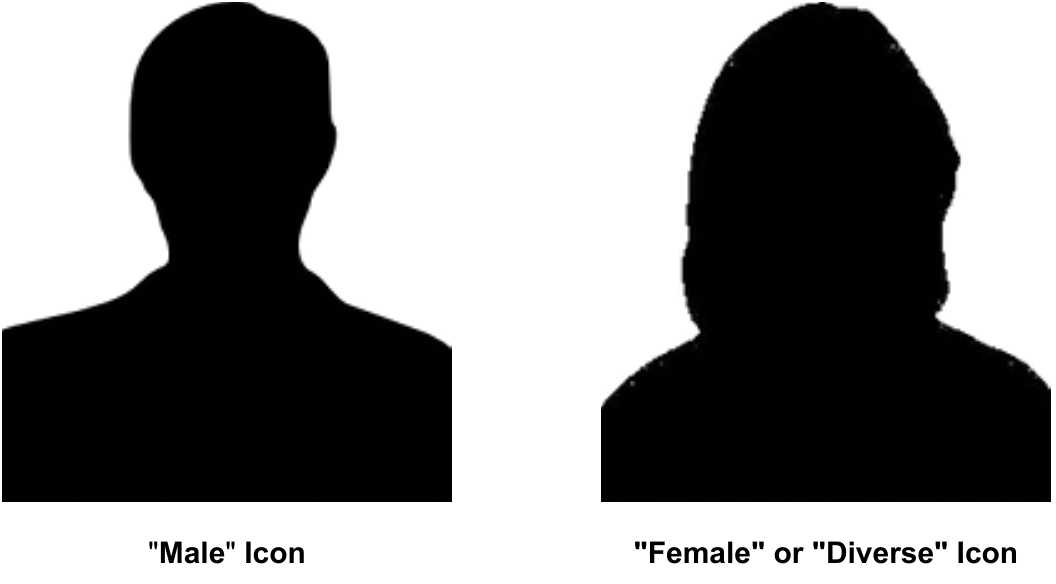}
    \Description[Figure with two silhouettes, one stereotypically male and one stereotypically female]{Figure with two silhouettes, one with short hair and wearing a collared shirt and one with long hair and rounded shoulders; text under the first silhouette says "male icon"; text under the second silhouette says "female or diverse icon" }
    \caption{An example of a bad application of user gender data: the website allows users to select one of three gender options (``male'', ``female'' or ``diverse'') which are used to determine the default profile picture for the website. Gender and profile picture should be uncoupled.}
    \label{fig:bad_gender_example}
\end{figure}

In order to use online services or other software programs, gender identification is often requested from users on sign-up. While this disclosure may be necessary for certain purposes, like providing medical care or helping people find community based on identity, gender is also collected for unclear or unessential reasons, such as advertising~\cite{Bivens_Haimson_2016}.  This presents privacy concerns for anyone who does not desire to share their gender with others online---such as the women who face harassment in video games~\cite{CoteVideoGames}. Additionally, unnecessary disclosure may also present issues of erasure and other trauma to transgender, non-binary or other gender non-conforming people.

Despite its universality in society, gender is not a single uniform concept. While identified in many cultures as a binary determined entirely by the two most prevalent sexual phenotypes, gender identity is described more accurately by the American Psychological Association as ``a person’s internal sense of being male, female or something else''~\cite{apatransgender}. Gender identity is distinct, although theorized to be closely related to both ``gender expression''---the way one presents or performs their gender to society---and one's gender assigned at birth, which is usually determined by physical characteristics~\cite{AmericanPsychologicalAssociation_2015}. Increased understanding and acceptance of those with marginalized gender identities has led to changes throughout society, such as an increased focus on bathrooms as one of the final realms of public gender segregation in the United States~\cite{Kopas_2012}. Online spaces have also introduced more inclusive gender fields, with Facebook now providing a free-form gender entry field rather than a fixed list~\cite{OKane_2015}.

Even with these promising changes, trans and non-binary individuals still face significant challenges in navigating computers and software, such as an inability to correctly gender themselves on digital forms and digital forms that do not allow them to opt-out of disclosure~\cite{10.1145/3461778.3462033}. Even when options are present for users to choose a gender which matches their identity, they may face erasure in the way that gender data is used. For example, a non-binary associate found that their employee profile at their place of work was accompanied by a silhouette which did not match their gender presentation because they had selected a gender which better matched their identity (see Figure~\ref{fig:bad_gender_example}). As they were not fully open about their non-binary identity with coworkers, the unanticipated use of gender data caused them privacy harm as well.  Prior work has also found that advertisers connected to sites with ``inclusive'' gender options often still put users into binary boxes for advertising purposes~\cite{Bivens_2017, Bivens_Haimson_2016}.

In order to contribute to the growing body of literature on how gender is treated in technological spaces and better understand why developers write ``shitty code''~\cite{Broussard_2019} with respect to gender, we present a first-of-its-kind study on developers and gender. Specifically in this paper we aim to answer the following research questions: 
\begin{itemize}
    \item \textbf{How do developers ask users for gender?}
    \item \textbf{How do developers use user gender data?}
    \item \textbf{What advice do developers receive about using gender data?}
\end{itemize}
To answer these research questions, we completed 15 open ended, recorded interviews with software developers, asking each questions about gender and gender disclosure. We also collected and categorized \numtotalposts~Reddit posts with a total of \numtotalcomments~comments. While distinct, these methods complement one another by providing alternative perspectives into the topic.

Through our analysis, we found that while most developers interviewed were receptive to inclusive gender forms, the code snippets present in our Reddit dataset overwhelmingly used binary gender disclosure forms with only ``male'' or ``female'' options. We also found that there was very little discussion of proper uses or treatment of user gender data on Reddit. Combined with our interviewees' seeming lack of previous exposure to considerations around how they should ask about or use gender data, this strongly suggests that developers have almost no exposure to inclusive practices with respect to gender. We ultimately recommend that more research should be done to collect a broader range of developer perspectives and confirm our initial findings. Additionally, we also suggest that software engineering and computer science educators consider highlighting research-backed, inclusive gender practices in courses to ensure that developers are exposed to best practices before they enter the workforce. 

The rest of the paper proceeds as follows: Section~\ref{sec:background} discusses necessary background information and other related work; Section~\ref{sec:methodology} describes in greater detail our research methodology for conducting both the interview study and our analysis of Reddit data; Sections~\ref{sec:results1} and~\ref{sec:results2} describe the results and key findings from our analysis of Reddit data and interviews respectively; Section~\ref{sec:discussion} discusses the broader implications of our findings and potential future work; and Section~\ref{sec:conclusion} concludes. 
\section{Background and Related Work}
\label{sec:background}

In this section, we discuss essential background research which informs our methodology. We start with a positionality statement to help readers understand our perspective and motivations for completing our research. We then move to discuss research from the field of queer studies, which grounds our methodology. Finally, we discuss the prior work in HCI which has analyzed gender, as well as previous work looking at the practices of developers. 

\subsection{Positionality Statement}
We acknowledge that our identities as researchers inform and shape the direction of our research~\cite{10.1145/3443686}. This paper draws on queer concepts of gender, and also discusses the impact of developers' treatment of gender on the experiences of transgender, non-binary, and other gender non-conforming people. The authors are cisgender members and allies of the LGBTQ+ community. The authors worked closely with trans and non-binary colleagues and friends, who gave feedback on our methodology and supported the project but opted not to participate as authors in favor of their own research projects. While we acknowledge that deeper insights can often be generated by those directly impacted by the identities being studied, we also acknowledge that engaging with somewhat toxic attitudes towards one's own identity can be tiring and traumatizing. We therefore welcome criticism of our endeavor, and hope to progress research that directly impacts our marginalized friends and colleagues from our privileged position as enthusiastic allies. While we pull heavily on research from the field of queer studies, our goal is not to theoretically advance the field, but rather to make an assessment of developers' knowledge, attitude and practices with regards to gender. We hope our research will positively impact how developers engage with gender so that queer people's marginalization in computing decreases.

\subsection{Gender, Sex, and Identity}

In the modern societies of North America and Europe, the dominant paradigm has held gender to be a strict binary determined by physical sexual characteristics.\footnote{This statement does not indicate that being transgender, non-binary, or otherwise not male or female is a new concept; non-binary conceptions of gender identity have existed across time and many different cultures~\cite{KesslerMcKenna1985}.} This ``cisgenderism''~\cite{doi:10.1080/19419899.2011.576696} is contrasted by the modern clinical and scholarly approaches to gender which accept not only gender but also sex as social constructs~\cite{KesslerMcKenna1985, Crawford_2000}. Not only is the conflation of sex with gender wrong, but researchers have identified a number of different ways sex anatomy can develop (intersex traits) which can cause individuals to fall beyond the traditional male or female sex boxes~\cite{Preves_2003, intersexterm}. Sex is therefore not an immutable property inherent to the universe, but rather a set of socially determined categories based on physical and genetic characteristics---usually male or female~\cite{KesslerMcKenna1985, West_Zimmerman1987, crawford10reappraisal, WestZimmerman2009}. 

Gender, by contrast, is a term which can be used to refer to several distinct but interrelated ideas:
\begin{itemize}
    \item \textbf{Gender Identity}: A person's inherent sense of being a man, woman, some combination of man or woman, or some other gender entirely~\cite{AmericanPsychologicalAssociation_2015, apatransgender}.
    \item \textbf{Gender Presentation}: The way in which one `does' gender through both their physical appearance and the way they act~\cite{KesslerMcKenna1985, crawford10reappraisal, West_Zimmerman1987, WestZimmerman2009}.
    \item \textbf{Gender Role}: The way in which people of a particular gender are expected to act by society \cite{KesslerMcKenna1985, crawford10reappraisal}.
    \item \textbf{Gender Assignment}: The gender one is assigned at birth~\cite{KesslerMcKenna1985, crawford10reappraisal}.
\end{itemize}
One may or may not present their gender in a way that makes others view them as a particular sex~\cite{West_Zimmerman1987}. These categories are highly contextual\footnote{For a simple example: one may publicly identify as the gender with which they were assigned at birth, but privately identify with a different gender identity.} and may vary over time with regards to a single individual~\cite{10.1145/3449113}. Like many other attributes, gender is not an isolated variable and its expression is influenced by other identities---such as disability~\cite{10.1145/3328320.3328369}. Within this framework, transgender people are those who transition from one gender to another~\cite{Haimson_2018, Stryker_2009}. This includes non-binary trans people---those trans people who do not identify with either male or female genders. There are also people who do not identify with the label trans but do identify as non-binary~\cite{Richards2016, LostyOConnor2018}.

Despite the scholarly recognition and increased public conversation of these concepts, trans and particularly non-binary gender identities are not well understood in many modern societies~\cite{Matsuno_Budge_2017, Garrison_2018}. Trans and non-binary people still face both overt and subtle discrimination from friends and family~\cite{Fuller_Riggs2018, Galupo_20141208}, others in public~\cite{lampe2020navigating}, and even medical providers ~\cite{Snelgrove_Jasudavisius_Rowe_Head_Bauer_2012, Romanelli_Lindsey_2020}. They also face larger structural barriers in attempting to live their gender~\cite{Shelton_Bond_2017}, like a lack of governmental recognition of their gender identity~\cite{Elias_Colvin_2020} or, in the world of research, the inability to change names on past research papers~\cite{Tanenbaum_2020}. Like many forms of discrimination, these issues are not uniform and are exacerbated by overlapping forms of oppression like racism and poverty, as Crenshaw first described in the concept of intersectionality~\cite{crenshaw1990mapping,VriesMattias2012, Longman_Marcellin_R_Bauer_I_Scheim_2013, Eastwood_Nace_Hirshfield_Birnbaum_2021, Balsam_Molina_Beadnell_Simoni_Walters_2011}.

In this paper, we focus our analysis of development practices on a broad category of discrimination against transgender and gender non-conforming people---erasure. Erasure is the systematic practice of ignoring or minimizing the existence of transgender and non-binary people~\cite{namaste2000invisible}. Erasure can have deep impacts on quality of life, as it contributes to the discrimination described above in myriad ways, such as through the denial of medical care~\cite{bauer2009don}. Even in seemingly mundane cases---like a customer feedback form which only includes male or female gender designations, excluding non-binary identities~\cite{Fonseca_2020}---such erasure adds to the many microaggressions which trans and non-binary people face~\cite{Nadal_Skolnik_Wong_2012} and which contribute to their increased rates of health problems~\cite{Balsam_Molina_Beadnell_Simoni_Walters_2011}. Addressing problems of erasure in software, as we do by analyzing developers behaviour and perceptions of gender, is therefore extremely important.

\subsection{Classification in Social Computing}
In the HCI community, especially in social computing, research on classification~\cite{bowker2000sorting} has long contributed to the scientific discourse: The origins, potential impacts and also potential harms on society are discussed both in the context of historical developments~\cite{bowker2000sorting, blackwell2017classification}, as well as in the context of qualitative coding, which, again, assigns categories and labels data~\cite{glaser2017discovery}. Notably, Bivens et al.~\cite{Bivens_Haimson_2016} examine how gender is treated on social media platforms, both from the perspective of a user and an advertiser. Using a walk-through approach, the authors elucidate how various design decisions act to integrate gender categorization into the platforms. They ultimately find that---even among platforms which do not directly collect user gender data or which include freeform gender fields---a binary conception of gender is ``baked in.'' They also convincingly argue that social media platforms' central role in the advertising industry gives them large influence over categorization practices throughout the entire computing ecosystem. Although we explore the problem from a different perspective, our work is motivated by the same problems as Bivens et al.~\cite{Bivens_Haimson_2016}, and our analysis is informed by their results. Our work, informed by prior work on classification and its impacts, considers the perspective of developers, who, whether or not they are aware of this, are performing classification work when implementing gender in software; we therefore contribute to and advance this conversation in CSCW.

\subsection{Gender and Computing}

Historically, work in the field of human computing interaction (HCI) surrounding gender has focused on traditional binary differences~\cite{Burtscher_Spiel_2020, Breslin_Wadhwa_2014,  Stumpf_Peters_Bardzell_Burnett_Busse_Cauchard_Churchill_2020} such as how gender differences impact people's use of problem solving software~\cite{Beckwith_Burnett_Wiedenbeck_Grigoreanu}. In line with the expanding definitions of gender accepted both in academia and public life, there is now an increasing body of research in HCI which looks specifically at the experiences of transgender, non-binary, and gender non-conforming people with technology. Prior work has looked at the ways that technology is used to create community for transgender and non-binary individuals and barriers they face to doing so~\cite{Scheuerman_Branham_Hamidi_2018, Haimson_Buss_Weinger_Starks_Gorrell_Baron_2020, Saha_Kim_Reddy_Carter_Sharma_Haimson_DeChoudhury_2019}; the problems caused by artificial intelligence's application to gender~\cite{keyes2018misgendering}, such as in facial recognition~\cite{scheuerman2019computers, scheuerman2020we} and automatic gender recognition~\cite{hamidi2018gender}; and the unique experience of gender transition and disclosure online~\cite{Haimson_Brubaker_Dombrowski_Hayes_2015, Haimson_2018, haimson2016constructing, haimson2016digital, 10.1145/3274332}, among other areas~\cite{10.1145/3301019.3323898, Haimson_Gorrell_Starks_Weinger_2020, Jaroszewski_Lottridge_Haimson_Quehl_2018}. While we do not directly investigate the problems which non-binary, transgender, and other gender non-conforming people face, such work is essential to understanding our project as it motivates our research. By understanding the approach that developers take towards gender, we hope to contribute to the reduction of such barriers to non-binary and transgender people's uses of technology.

There has also been some research which looks specifically at web forms. Most notably, Scheuerman et al.~\cite{Scheuerman_Jiang_Spiel_Brubaker_2021} present an evaluation of various gendered web forms with 350 non-binary participants, finding that forms which only offer male/female options are uncomfortable to participants. The paper, while not specifically studying developers as we do, recommends that developers consider alternatives to asking about gender---such as instead asking about pronouns---or offer inclusive gender options if gender is indeed necessary~\cite{Scheuerman_Jiang_Spiel_Brubaker_2021}. We build upon this research by using the paper's insights to evaluate developers' use of gender. We were also motivated by the experience of the queer researcher Dr. Katta Spiel, who has discussed their experience with technological infrastructure and gendered forms in several previous publications~\cite{10.1145/3461778.3462033, 10.1145/3290607.3310425}. Their recommendations relating to survey construction and gender---written with two other researchers---also informed our screening survey form construction~\cite{10.1145/3338283}.

\subsection{Developers and Reddit}

One final area of related work which is important to discuss is prior work which has looked at the behaviour of developers as well as those studies which have used Reddit in particular. Due to their essential role in creating software, programmers and other developers have long been studied in computing research~\cite{rasch1992factors, 799955}. Researchers have researched both the practices of developers---like where they log~\cite{fu2014developers} and how they test applications~\cite{8094467}---as well as their perspectives on specialized topics, like privacy~\cite{hadar2018privacy}, security~\cite{acar2017developers}, and usability~\cite{hoegh2006usability}. To the best of our knowledge, we are the first to look specifically at practices and perspectives with respect to gender. 

Researchers have specifically analyzed the social media website Reddit as a community of practice for  developers~\cite{Kou_Gray_Toombs_Adams_2018, 10.1145/3283827}. Reddit is a news aggregation and discussion forum that was founded in 2005. One of the key features of the website is the sub-forums called ``subreddits'' which are almost all community-created and moderated. Users can post either links or ``discussion'' posts on a subreddit. Each post is accompanied by a comment section in which others can reply to the post. Both posts and comments include voting buttons, allowing users to ``upvote'' or ``downvote'' posts to affect the order in which both posts and comments are displayed to others. An upvote increases a post or comment's score, while a downvote decreases a post or comment's score. Reddit, thus, presents a rich source of both quantitative and qualitative data about users' perspectives.~\cite{Medvedev_Lambiotte_Delvenne_2019} 

There are a number of communities centered around learning or discussing programming, with over 3.5 million users subscribing to the ``programming'' subreddit\footnote{www.reddit.com/r/programming} alone~\cite{frontpagemetrics, Hardin_Berland_2016}. While there are other websites like StackOverflow\footnote{www.stackoverflow.com} which are more directly about programming, Reddit's discussion-oriented format makes it a rich repository for qualitative perspectives on programming. Researchers have used Reddit~\cite{li2021developers} and StackOverflow data~\cite{yang2016security, lopez2019anatomy} to gain insights about how developers think about specific topics, such as privacy~\cite{li2021developers} and security~\cite{yang2016security, lopez2019anatomy}. We contribute to this larger body of work by exploring how developers discuss gender on Reddit.

\section{Methodology}
\label{sec:methodology}
In this section, we explain our methodology, beginning first with a description of our collection and analysis of programming sub-forums on Reddit before describing our interview process and analysis. While described and analyzed separately, these methods complement one another in the following ways: participants in an interview study may, intentionally or not, misrepresent their true opinions or practices. Reddit is a pseudo-anonymous space wherein users may feel more comfortable expressing their true opinions. Reddit also reflects a subset of the public conversation and practice around gender, which was not assessed in our interviews. Similarly, in interviewing developers, we were able to directly interrogate issues which were very infrequently discussed in our Reddit data. We believe both of these studies provide valuable insight.

\subsection{Reddit Study}

The first part of our study consists of qualitative analysis of posts on programming subreddits on Reddit, with~\numtotalposts~posts and~\numtotalcomments~comments collected on June 22nd, 2021 using Reddit search, seeded with terms related to gender. By analyzing posts on Reddit, we are able to get insight into how programmers speak to one another about gender data outside of the formal environment of an interview. We are also able to more quantitatively evaluate the various ways that developers use gender data in applications. 

\subsubsection{Data Collection} The first step in the data collection process was to select subreddits for analysis. Criteria for subreddits to be included were broad: first, they had to be focused on programming or user experience/user interface design; second, they could not be focused on a specific programming language as our focus is on general software development practices, rather than any particular language. Additionally, we felt that subreddits focused on a particular language may be less likely to include discussion on gender. From those subreddits which met these criteria, we selected 11 covering a broad range of topics from web development (/r/webdev and /r/web\_design) to game development (/r/gamedev). The full list of subreddits used---and the number of posts collected from each---is in Table~\ref{tab:subreddits}.
\label{app:reddit_data}
\begin{table}
\caption{Subreddits used for data collection, with number of subscribers on date of data collection, number of posts collected, and age of subreddit as reported by Reddit.}
\label{tab:subreddits}
    \begin{tabular}{@{}lccc@{}}
        \toprule
        \textbf{Name} & \textbf{\# of Subscribers }& \textbf{Community Age (years)} & \textbf{\# of Posts Collected}\\ 
        \midrule
        /r/programming & 3,500,000 & 15 & 84 \\
        /r/learnprogramming & 2,300,000 & 11 & 422 \\
        /r/compsci & 1,500,000 & 13 & 29 \\
        /r/webdev & 880,000 & 12 & 91 \\
        /r/gamedev & 570,000 & 13 & 167 \\
        /r/web\_design & 550,000 & 13 & 39 \\
        /r/coding & 290,000 & 11 & 6 \\
        /r/androiddev & 180,000 & 12 & 20  \\
        /r/iOSProgramming & 100,000 & 10 & 38 \\
        /r/userexperience & 81,000 & 13 & 15 \\
        /r/UI\_Design & 52,000 & 8 & 6 \\
        \bottomrule
    \end{tabular}
\end{table}

In the remainder of the paper, relative size comparisons for subreddits is based on the number of subscribers listed in this table. After selecting subreddits to investigate, we then created a list of search terms which we felt would produce germane results. This included general terms related to gender---like ``gender'' and ``sex''---as well as terms related specifically to marginalized groups---like ``LGBTQ'' and ``transgender.'' The full list of search terms is in Table~\ref{tab:search_terms}. Reddit posts were collected using the Reddit search API\footnote{\url{https://www.reddit.com/dev/api/}} seeded with relevant search terms and limited to one of the programming/design related subreddit at a time. While using the Reddit search API may exclude some user- or moderator-deleted posts which are preserved in other archives, this method of data collection has the benefit of mimicking more accurately how an average user would use the website. The content, metadata, and comments for each post were then collected and downloaded on June 22nd, 2021. Using this method, \numtotalposts~posts and \numtotalcomments~comments were collected.

\begin{table}
\caption{Search terms used for data collection with number of posts collected.}
\label{tab:search_terms}
    \begin{tabular}{@{}lc@{}}
    \toprule
    Search Terms & Number of Posts \\ 
    \midrule
    ``gender'' & 528 \\
    ``non-binary'' and ``nonbinary'' & 172 \\
    ``transgender,'' ``trans,'' and ``trans gender''  & 147 \\
    ``LGBTQAI,'' ``LGBTQ,'' ``LGBT'' & 46 \\
    ``pronouns'' & 24 \\ 
    ``sex'' & 0 \\
    \bottomrule
    \end{tabular}
\end{table}

\subsubsection{Analysis}

Prior to further analysis, it was necessary to evaluate posts, as many were not germane to gender. The collected threads were placed into three categories:
\begin{itemize}
    \item \textbf{Relevant Discussion}: Threads which explicitly give or request advice on gender disclosure or the appropriate way to use/implement gender data.
    \item \textbf{Using Gender}: Threads which were not about gender disclosure but included code or other descriptions of using gender in software; we further analyzed such threads for the representation of gender used in the program. (e.g. binary, binary with an ``other option,'' free entry, etc.)
    \item \textbf{Irrelevant Discussion}: Threads which were not about user gender data at all, such as those which only used a particular term in passing; this category includes discussions of gender identity in fictional media like video games.
\end{itemize}
These categories allow us to investigate all three of our research questions. Posts in the ``relevant discussion'' category reflect the advice developers receive on using gender data, while the ``using gender'' category gives insight into how developers integrate gender into software. Categorization was completed by the primary author in consultation with the other authors.

After categorization, we calculated summary statistics on Reddit posts in each group, including median ``upvote ratio'' (the ratio of upvotes to downvotes on a post), median number of comments on each post, and the median number of upvotes. The upvote scores are not exact, as Reddit does not provide the number of downvotes or the exact upvote ratios for posts to avoid vote manipulation. We also conducted in-depth qualitative analysis of the significant posts which discussed, requested, or gave advice on proper practices for handling gender. The complete results of our analysis are in Section~\ref{sec:results1}.

\subsubsection{Limitations}

Our Reddit procedure has some limitations. First, as mentioned above, our sample is intentionally biased and cannot be said to reflect the general rate or nature of all conversations on user gender data on programming subreddits; this bias is necessary to analyze posts on gender data which are likely not common. Second, due to the use of Reddit search to collect data, our results may not be exactly replicable; we plan to preserve our data for 10 years and will provide it to interested researchers.\footnote{While our data is technically public, some of our search terms relate to topics which can be sensitive. We do not want to share our raw data publicly out of an abundance of caution.} Third, the level of anonymity afforded by Reddit may affect the opinions users express or the manner in which they express their opinions. Finally, while we have attempted to clearly define our analysis categories, there is inherent subjectivity which cannot be avoided. Ultimately, despite these limitations, we hope to provide valuable initial insight into discussion of gender data among developers.

\subsection{Interview Study}
\label{ssec:interview_method}

In addition to our Reddit study, between June and August of 2021, we conducted a semi-structured interview study of 15 developers, most of whom had worked with gender in software. These interviews allowed us to ask developers directly about user gender data and gain more insight into their thinking around gender. The interviews were typically around 20 minutes in length and conducted in an open-ended fashion. As much as possible, the interviewer avoided prompting interviewees, instead allowing them to speak on topics in their own words. 

\subsubsection{Participant Recruitment}

Our eligibility criteria were broad in order to capture a wide spectrum of experience from recent graduates of engineering schools to those who have worked in software development for years. Interviewees were required to be at least 18 years old, comfortable with conducting the interview in English, and they needed to have at least a year of professional programming experience or a degree in a relevant field---such as computer science. We established eligibility via a pre-screener.

Participants for interviews were recruited in two main ways. First, we posted an advertisement on the freelancing platform UpWork.\footnote{\url{https://www.upwork.com}} This advertisement included short screening questions based on our eligibility criteria to prevent spam or ineligible participants. All except one person who responded to the posting were eligible. Using UpWork's tool, which allows users to directly invite freelancers suggested by the platform, we also sent the advertisement directly to several freelancers in the United States in order to diversify the geography of the interview pool\footnote{Prior to directly recruiting participants from the United States, none of the participants in our UpWork interview pool were from the United States. Due to our focus on English language software development practices, we felt that participants from this area were important to more actively recruit.} and increase the number of participants. Interviewees were also recruited using an advertisement on the social media platform Twitter. The advertisement was posted as an image by one of the authors on their public Twitter account. The Twitter account had at least 200 followers\footnote{These followers were predominantly made up of professional and personal contacts of one of the authors, including a relatively large number of recent college graduates and other computing researchers.} at the time that the tweet was posted. The tweet received over 7,000 impressions during the study, according to Twitter's analytics.  Ultimately, \numupworkinterviews~participants were recruited through UpWork and \numtwitterinterviews~participants were recruited through Twitter.

One of our interviewees was recruited in a distinct manner. In order to ensure that we got the perspective of a developer who had worked with gender and marginalized groups, we reached out directly to a programmer who had previously worked on a website for transgender and gender-non-conforming people. This participant was asked the same questions as other participants, but their unique insights are discussed separately when appropriate. 

We interviewed until we reached theoretical saturation; that is, no new ideas were discussed in the interviews~\cite{corbin2014basics}. All interviews were conducted by the same author---a man in his 20s.

\subsubsection{Interviewing Participants}
Once participants were shown and digitally signed an informed consent form, they completed a pre-interview survey about their demographic and professional background to better describe the diversity of our sample. This survey also allowed us to focus on developers' qualitative experiences in interviews. The complete survey can be found in Appendix~\ref{app:screener_survey}. In the interviews, we were interested in our participants' professional experiences with gender forms/requests/coding in software development, and we developed our semi-structured interview guide to reflect this. We discussed our interview guide with one non-binary developer who is also actively engaged in LGBTQ+ community building. Finally, we conducted a pilot interview with another researcher to confirm both the length of the interview and that questions were appropriate. The pilot interview is not included in the results. The full interview guide can be found in Appendix~\ref{app:interviewprocedure}.

Our interview guide consists of three parts: pre-interview procedure, interview questions, and example gender options. The pre-interview procedure included instructions on how to introduce oneself, how to give an overview of interview procedure, and other reminders. While only one author conducted all interviews, having a defined pre-interview procedure helped ensure that interviews were performed consistently. Once the interview began, we moved to the second part of the interview guide which includes a list of questions. The exact phrasing of questions varied slightly between interviews. The interviewer also occasionally asked follow-up questions for clarification. Our initial questions discuss general and gender-aware programming experience. We then move to discuss how participants would hypothetically handle gender data, including asking about gender, whether they consider gender private, and finally whether they have any concerns about using gender data.

After all other questions were answered, we then displayed example gender disclosure options contained in the third part of the interview guide. We discussed each with the interviewees to get their perspective on whether each option is appropriate and when it may be useful. We first showed them the least inclusive gender disclosure prompt, which included only ``male,'' ``female,'' and ``prefer not to answer'' options (the ``binary'' option). We then showed a disclosure prompt which included the same options as the previous example as well as a third option of ``diverse,'' as is used in government forms in Germany~\cite{germany_gender} (the ``ternary'' option). Finally, we showed them our most inclusive option, which was developed based on the recommendations in Scheuerman et al.~\cite{Scheuerman_Jiang_Spiel_Brubaker_2021} as well as other works discussed in Section~\ref{sec:background}. This example identified the data being asked for as ``gender identity'' rather than ``gender,'' as was asked in the previous questions. It included the options ``man,'' ``woman,'' ``non-binary,'' and ``self-identify'' with a text box. We finally showed them the current ISO/IEC 5218~\footnote{\url{https://www.iso.org/standard/36266.html}} standard for storing human sex: 0= not known; 1= male; 2 = female; 9 = not applicable.\footnote{While the standard is not technically meant to be used to gender, pre-interview research found that the standard was offered in forum posts around the internet.} See Appendix~\ref{app:interviewprocedure} for the full texts and options of all variations. 
Finally, we concluded the interview by asking if participants had anything else they wanted to share or discuss. After stopping the recording, we also asked if they had any questions for the interviewer before the interview call was terminated.

\subsubsection{Ethics \& Data Protection}
Prior to participation in the interview study, participants consented to participation in the study as well as audio recording via our consent form, in which we listed our data collection, storage and use practices. We informed them about their right to withdraw from the study at any time without any repercussions or loss of benefits, as well as the option to skip any questions in the interview. We also agreed to quote interview participants in a non-identifiable manner. While this may limit the quotes or descriptions we can report, it also allowed interviewees to speak freely about their experiences. We use quotes throughout Section~\ref{sec:results2} to enhance our analysis. Finally, we answered participants' questions about procedure and the purpose of the study, and debriefed after the interviews.

All interviews were conducted using the Zoom\footnote{\url{https://zoom.us/}} meeting service. Meetings were recorded via Zoom's built-in recording feature. While the interviewer had video conferencing enabled for all interviews, participants were free to enable or disable video as desired. Visual aspects of interviews were not analyzed. Audio-only recordings were then stored in a secure, VPN-accessible institutional cloud and transcribed by a GDPR-compliant service. Transcripts were pseudonymized and also stored in the secure cloud, both for storage and for collaborative coding. Participant data required for payments was stored separately and only used for payments. Participants were paid US $\$30$, as we expected interviews to last up to 30 minutes, and wanted to pay US $\$60/$hour, the average rate for Upwork developers. In practice, the interviews were shorter (mean = 18.0 minutes, std = 3.24 minutes).
The study was approved by our institution's ethical review board and data protection office. 

\subsubsection{Interview Data Analysis}
Once all interviews were complete, we used deductive and inductive coding for our data analysis~\cite{mcdonald2019reliability, corbin2014basics}. We began with an \textit{a priori} code list that corresponded to our interview and research questions. Following this, the first author developed a codebook by inductively going through the interviews, continuously checking in and discussing with the team, and operationalizing the codebook by formally defining the codes. Once the codebook was stable, the first author coded all transcripts in coordination with the research team. The codebook---including definitions---can be found in Appendix~\ref{app:codebook}. 
Throughout and following the coding process, we met and discussed emerging concepts, themes, and relationships, which allowed us to explore how developers relate to gender when programming. In Section~\ref{sec:results1}, we discuss the results of coding as well as other significant observations we made.

\subsubsection{Limitations}
Like any interview study, our participants' responses are subject to biases, including recall, self-report and social desirability bias. In addition, we recruited developers from Twitter and Upwork. Demographically, our sample skews young and towards men, which is sadly reflective of the general state of gender representation among developers.\footnote{See \url{https://insights.stackoverflow.com/survey/}} While we do not claim a representative sample, we did try to mitigate recruitment biases by (a) recruiting specifically from within and outside the United States; (b) recruiting with and without use of our own social networks; (c) recruiting from within and outside the ``progressive, gender-inclusive bubble'' that the authors mostly live in. Since the smaller sample of qualitative participants does not support generalizability, we did not perform testing for differences based on either demographics or recruitment channels.
Especially in contrast to the Reddit data,  the presence of an interviewer may have affected the interview results. Interviewees may have been less willing to express a preference for noninclusive gender options when speaking with a researcher as opposed to those posting on an anonymous forum (i.e. social desirability bias). Additionally, developers who agree to participate in a study explicitly about gender practices may be more inclined towards inclusive practices. Indeed, one participant shared---after the interview session---that they responded to our Upwork ad because of their interest in accessibility and inclusively. 
While these biases may have influenced participants' stance towards gender, they did report noninclusive practices, and we think that our interview results give meaningful insights into how developers program gender.


\section{Reddit Results}
\label{sec:results1}

In this section, we present the results of our study of discussion of gender on Reddit. We start by describing our dataset in broad terms before moving to discuss specific aspects of the data. 

\subsection{Overview}

\begin{table}
    \centering
    \caption{Overview of post statistics in the Reddit dataset.}
    \begin{tabular}{@{}lcccc@{}}
    \toprule
     & & \multicolumn{3}{c}{\textbf{Median}} \\
    \cmidrule(lr){3-5}
    \textbf{Category} & \textbf{Number of Posts} & \textbf{Upvote Ratio} & \textbf{\# Comments}  & \# \textbf{Upvotes}\\
    \midrule
    Relevant Discussion & \numrelevantposts & 0.57 & 5 & 1 \\
    \midrule
    Using Gender &  \numusinggender & 0.81  & 4 & 1 \\ 
    \midrule 
    Irrelevant Discussion & \numirrelevantposts & 0.76 & 5 & 2 \\
    \midrule
    Total & \numtotalposts & 0.75 & 5 & 1 \\
    \bottomrule
    \end{tabular}
    \label{tab:reddit_overview}
\end{table}

 The vast majority of posts in our dataset were irrelevant to our research questions (see Table~\ref{tab:reddit_overview}). Very few threads discussed appropriate practices towards user gender. Out of the~\numtotalposts~total threads, only~\numrelevantposts~(1.7\%) either offered or requested advice on how to use gender data in software. The median upvote ratio of those posts was 0.57, which is much lower than the overall median.~\numusinggender~posts discussed topics other than gender data but did include code or other descriptions of using gender data in a program.

\begin{figure}
    \centering
    \caption{Number of posts from each year in the Reddit dataset.}
    \includegraphics[width=\textwidth]{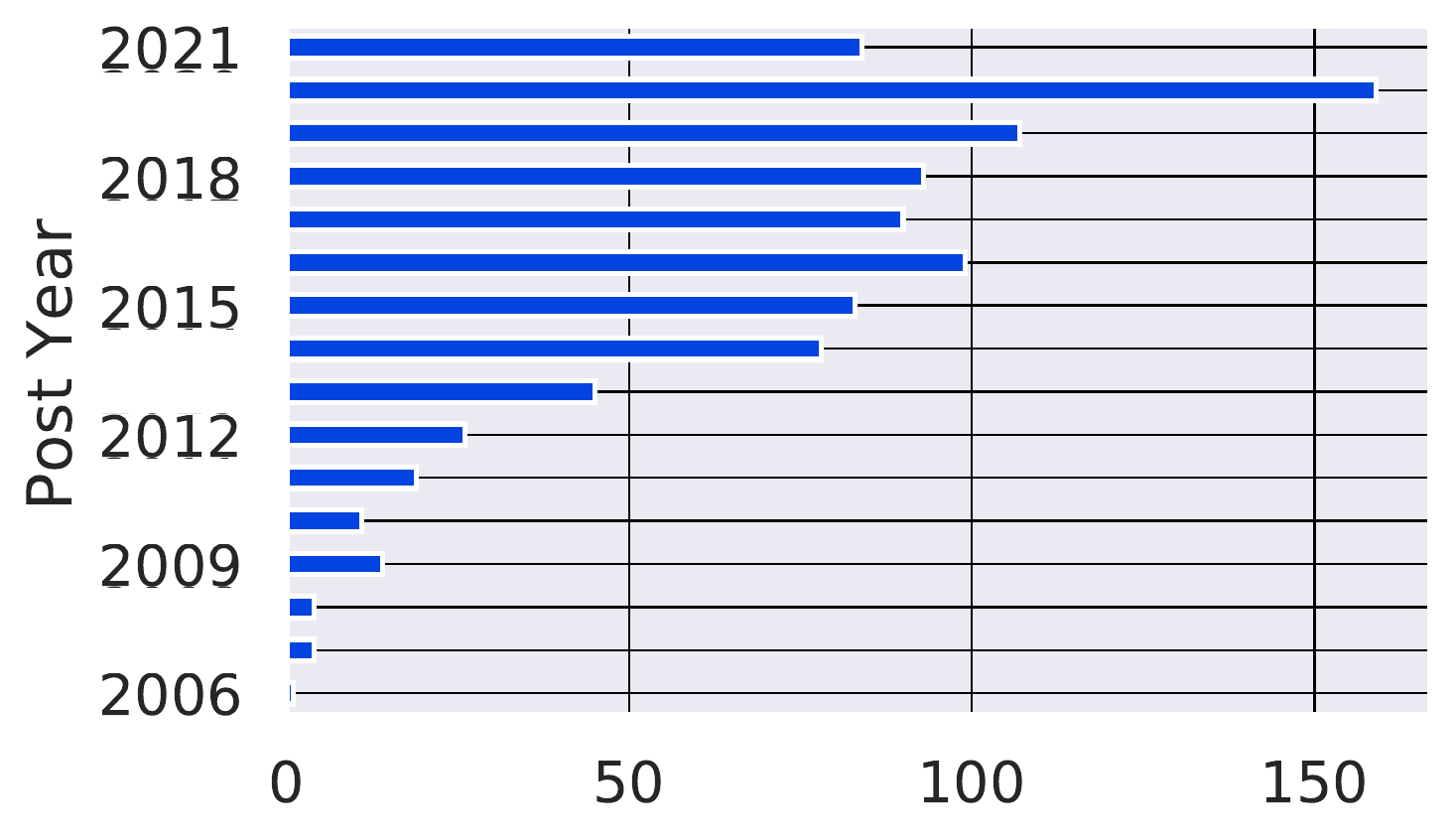}
    \Description[A horizontal bar plot showing that most posts are from after 2014]{A horizontal bar plot showing that there are less than 25 posts in each year from 2006 to 2010; between 25 and 50 posts in each year from 2012 to 2013; between 50 and 125 posts in each year from 2014 to 2019 and 2021; and finally, there are over 150 posts from 2020}
    \label{fig:reddit_time_graph}
\end{figure}

As shown in Figure~\ref{fig:reddit_time_graph}, most posts are from within the last 5 years. 58\% (533) of posts in the dataset are from 2017 or after and only 5.8\% (53) of posts in the data set are from 2011 or before. This result is in line with the exponential growth in Reddit posts over time reported by Medvedev et al.~\cite{Medvedev_Lambiotte_Delvenne_2019}. This observation may also be attributable to an increase in conversations on gender over time, but we do not evaluate this assertion. Regardless, this does indicate that our dataset is mostly made up of recent content rather than older posts which may not reflect the current state of discourse. 

Before moving to discuss the two main categories of post in detail, we briefly want to touch on the ``irrelevant'' category. We did not perform a systematic evaluation of irrelevant posts; however, during categorization, we did notice that much of the content in the category was about gender diversity and inclusion in the field of software development. While analyzing such conversations is beyond the scope of our paper, the large number of such threads may be a reflection of the increased focus on gender diversity in contemporary life and particularly in technical fields like software development~\cite{Huang_Krivkovich_Starikova_Yee_Zanoschi_2019}.

\subsection{Direct Discussion of Gender Data}

As shown in the previous subsection, we found very little content which discussed appropriate practices with regards to user gender data. Of the 16 posts on the topic, half were links to external websites with discussion on using gender data (``link posts''). The remaining eight posts were self-contained discussions about gender data (``discussion posts''). The majority of the posts in both subcategories spawned very little conversation. Two of the posts received no comments. 11 posts received between 0 and 10 comments, and three posts received 20 comments or more.  12 (75\%) of the posts were made after 2017 while only one of the posts were from before 2011 (6.3\%), roughly mirroring the overall dataset's bias towards more recent posts. 

The post which spawned the most discussion was a link post on the /r/programming subreddit. The post focused on Google Cloud Vision's choice to no longer return gendered labels like ``man'' or ``woman,'' and it had over 480 comments. It also had the highest score of the subset of relevant posts with 130 upvotes (0.72 upvote ratio). The post consisted of a link to Google's AI ethics principles, including discussion of avoiding gender bias in programming AI~\cite{OurPrinciples}. The post title copied language directly from Google and seemed neutral towards the Google guidelines. In contrast, the vast majority of comments were against removing gendered labels from the API, with the most highly voted comments suggesting that ``lunatics'' or ``inexperienced college grads'' have taken over Google. While some comments expressed positive sentiment towards removing binary conceptions of gender from the API, they almost all had negative voting scores or were marked as ``controversial'' by Reddit---indicating that despite having a positive score, they received many negative votes.

Only two other link posts had a positive score. One from the /r/coding subreddit, with a 0.63 upvote ratio, suggested that developers should use functions to easily program custom pronouns based on user gender. It recommended the use of singular ``they/them'' pronouns as a default option. Comment feedback was largely positive. There were a few comments which expressed light criticism, stating that the blog post was trivial. Users also pointed out that the proposed function only worked for English. This post only very lightly touches on conceptions of gender, which may have helped it avoid the harsh reaction levied towards other threads. 

The other positively received link post---from the /r/userexperience subreddit---led to an article on creating more inclusive gender options and had an 0.70 upvote ratio. The advice in the article generally mirrors that of Scheuerman et al.~\cite{Scheuerman_Jiang_Spiel_Brubaker_2021}, recommending that programmers avoid asking for gender when possible, give a good reason to ask about gender when it is necessary, and include inclusive options in all gender forms, among other advice. It may be undercut, however, by the only reply stating that the ``gender diaspora'' has gotten ``crazy.''  

All of the other link posts have a 0.50 upvote ratio or below, indicating that such posts are not well-liked by the programming communities on Reddit. The least well-received post---with an upvote ratio of 0.22---was a link to an abstract discussion of the construction of gender for artificial intelligences, which received no comments. Of those posts which do have comments, the criticism seemed to mirror that present under the post on Google Cloud Vision: dismissal of the topic as unworthy of discussion. 

Discussion posts in the dataset were largely confined to the smaller subreddits in our corpus, including /r/userexperience, /r/UI\_Design and /r/web\_design. All had scores of 0.50 or above, and all except one received comments. Discussion posts covered a wide variety of topics, from how to tailor user experience to particular gendered audiences to whether the term ``sex'' or ``gender'' was most appropriate for a particular form. While the variety of posts and small sample size means that it is difficult to generalize, the advice was largely inclusive. For example, the most highly voted response to a discussion post which inquired about the best practices for requesting user gender gave advice similar to the best practices described by Scheuerman et al.~\cite{Scheuerman_Jiang_Spiel_Brubaker_2021}. The most germane critique of the advice users were given in response to their questions is that, as one might expect from a casual internet forum, advice was not sourced and relied heavily on personal experience. 

Largely, then, Reddit users looking for advice on gender seem to receive inclusive advice when they request it through discussion posts. On the flip side, link posts about gender, particularly on the larger subreddits in our
corpus, seem to receive dismissive or, at worst, extremely negative reactions. One reason for this disparity may be the different topics of the subreddits. For example, the /r/UI\_Design and /r/userexperience subreddits are focused on the human aspects of software development in contrast to the broad technical focus of larger subreddits like /r/programming or /r/compsci. While we do not have any data on the actual audiences of the communities, we believe that this topical difference may have fostered a community which was better able to talk about gender. 

The most concrete takeaway from this data, however, is that very little discussion on best practices towards gender occurs on Reddit programming communities. While our inability to find posts on handling gender data is not absolute proof that these conversations are not happening somewhere on Reddit, they seem to not be occurring on the larger subreddits in our corpus. 

\subsection{Other Uses of Gender Data}

\begin{table}
\parbox{.45\linewidth}{
    \caption{Use cases of gender in the Reddit dataset.}
    \centering
    \begin{tabular}{@{}lc@{}}
        \toprule
         \textbf{Use} & \textbf{Frequency} \\
         \midrule
         Unspecified & 142 \\
         \midrule
         Audience Demographics & 35 \\
         \midrule
         AI/ML & 19 \\
         \midrule
         Health/Fitness & 17 \\
         \midrule
         Healthcare & 8 \\
         \midrule
         Dating & 5 \\
         \midrule 
         Other (Specified) & 20 \\
         \bottomrule
    \end{tabular}
    \label{tab:uses_of_gender}
}
\parbox[]{.45\linewidth}{
    \caption{Types of gender fields in the Reddit dataset.}
    \centering
    \begin{tabular}{@{}lc@{}}
        \toprule
         \textbf{Representation of Gender} & \textbf{Frequency} \\
         \midrule
         Unknown & 149  \\
         \midrule
         Binary & 80 \\
         \midrule
         Binary with third option & 9 \\
         \midrule
         Facebook/Instagram & 5 \\
         \bottomrule
    \end{tabular}
    \label{tab:type_of_gender}
}
\end{table}

While our main aim in analyzing Reddit was to observe the advice that programmers give one another, we were also able to use our dataset to look at some of the ways that programmers are using gender. As our sample is intentionally biased towards particular keywords, we cannot assume that the statistics reflect the broader nature of all programs---or even all code posted on Reddit. However, our analysis contributes to the larger view of how programmers use and view gender.

Table~\ref{tab:uses_of_gender} shows the use cases of gender data found in our dataset, with more than 5 occurrences in the dataset. The vast majority of posts (142) did not specify a use for gender data. Of the posts which did specify a purpose for collecting gender, the most common use cases were audience demographics (35)---such as for advertising or developing a business plan---and artificial intelligence or machine learning (19)---such as classifying photos. As the posts used to collect this data were not explicitly about gender, we cannot necessarily infer anything from the large number of threads which do not explain why they collect gender data. It is notable, however, that the most common specified use case is not directly necessary for the functioning of an application: demographic analysis.

The other data point collected when categorizing posts was the representation of gender used in the application, such as binary or binary with a third option. Similar to categorizing the uses of gender data, most posts (149) did not contain enough information to determine the representation of gender. Some of the ``unknown'' posts did include information about the type of variable used to store gender information, with 22 specifying gender as a ``string'' type variable and 16 specifying gender as a ``char'' type variable; we cannot infer from this how exactly the program implements gender. Overwhelmingly, the threads which include information on gender ``type'' featured binary conceptions of gender, either by storing gender as a ``boolean'' type or by only including ``male'' or ``female'' options in their code. A small number of posts (nine) included an ``other'' option along with the``male'' and ``female'' options. We also saw some posts (five) which used gender values imported from Instagram or Facebook. Ultimately, then, most of the Reddit posts which feature programs which use gender data---and include enough detail to identify an approach to gender---use noninclusive, binary options. These data, particularly in light of the hostility observed in direct discussions of inclusive approaches to gender information, suggest that the dominant conception of gender held by developers on these subreddits is essentially binary. That is to say, they likely view gender as consisting of two opposite categories---"male" or "female"---which are universal.

One important caveat to this finding is that not all of the posts are real world examples. Many posts explicitly mentioned that code excerpts were from school projects.\footnote{Several posts from different users appeared to reference the exact same homework exercise on programming a BMI calculator.} Additionally, the large prevalence of posts from the subreddit /r/learnprogramming in the dataset means that even non-school projects may still be learning exercises. These data are still valuable, however, as such posts still reflect the way that people learn how to use gender data in programs. Further, early in a software programmer’s education, school projects might be the most appropriate location to broach the subject of gender.

In summary, we observed very little direct discussion about gender in our Reddit dataset. In the few posts where users requested advice on how to treat gender data, advice was generally inclusive. Unfortunately, general discussion about inclusive concepts of gender was faced with hostility. Moreover, binary concepts of gender were the most common observed in practice on Reddit. With these findings in mind, we now move to look at the results of our interview study.

\section{Interview Results}
\label{sec:results2}

In this section, we present the results of our interview study. We start by discussing the results of coding before moving to qualitatively discuss other significant findings. When relevant, we specifically highlight the views of the participant who had previously worked on a social media community for transgender, non-binary, and other gender non-conforming people. While this participant's experience is likely not typical, we highlight it to elevate their unique perspective on developing for marginalized communities. This interviewee will henceforth be referenced as the ``expert participant.'' At points, we compare our findings to those from the previous section for illustrative purposes. Due to the different methodologies, however, we do not draw definite conclusions about how participants may differ from users in our Reddit sample. 

\subsection{Interviewee Demographics}

Using the methods described in the previous section, we recruited 15 participants. Eight were from the United States, two were from Germany, two were from Pakistan, and three were from other countries. Participants ranged from 19 to 39 years old (median = 23). Twelve were men, two were women, and one participant chose the option to self-describe, but did not fill in the text box. Six were white or of European descent, four were South Asian, two were Hispanic or Latino/a/x, one was Middle Eastern and one was Black or of African descent. Participants had educational backgrounds ranging from being a high school graduate to having a master's degree, with six participants reporting a bachelor's degree, three reporting a master's degree, two reporting some time at college, and two reporting a high school diploma. Of those with degrees, seven reported having a degree related to computing, such as computer science or engineering, while one reported having a degree in the social sciences. Finally, participants reported one to ten years of development experience (median = four years).

\subsection{Coding Results}

Our interviewees were roughly evenly divided in terms of development background. Four participants (26\%) identified themselves as primarily frontend developers, five participants (33\%) described themselves as backend developers, and five participants (33\%) described themselves as working on both backend and frontend development and were coded as fullstack developers. One participant did not give a clear answer to the question. Almost all interviewees had experience on projects working with both personal data and gender data: 14 participants (93\%) reported experience with user personal data and 11 participants reported previous experience with gender data (73\%). 

As Table~\ref{tab:interview_uses_of_gender} shows, interviewees brought up many similar applications of gender data to those observed in the Reddit dataset. Participants brought up uses of gender data throughout the interviews, but most commonly after they were asked about their experience using gender data in programming applications. Demographic analysis was the most frequent (six times) use mentioned by our interviewees and it also appeared most frequently in our Reddit dataset. While dating was frequently mentioned (four times) by interviewees as a use for gender information, none had worked on a dating application before. This use could indicate a conflation of gender with sexuality, but we find it more probable that this expresses the assumption that gender is an essential part of dating. Interviewees also discussed using gender in other ways where it is likely not essential. Legal/financial purposes and AI/ML came up in multiple interviews. In the case of legal/financial purposes, one interviewee mentioned that they thought gender was required by regulation: ``if you sign up for contracts with some companies, most times you need gender data for verification. The project I worked on before was for a ... telecommunications company... and they sure collected gender data when someone signed up there.''  We were not able to verify what regulation he was referencing, but his statement seems to suggest that he believed it was required because it was collected. There may not have been an actual legal requirement.

\begin{table}
    \caption{Uses of gender discussed in interviews; frequency represents number of interviews in which the use case was brought up.}
    \begin{tabular}{@{}lc@{}}
        \toprule
        \textbf{Use of Gender Code} & \textbf{Frequency} \\ \midrule
        Audience/Client Demographics &  6\\ \midrule
        Dating & 4 \\ \midrule
        Healthcare & 3 \\ \midrule
        Diversity Protection & 3 \\ \midrule
        Legal/Financial Purposes & 3 \\ \midrule
        No Specific Purpose & 3 \\ \midrule
        AI/ML & 2 \\ \midrule
        3rd Party Telemetry/Advertising & 1\\
        \bottomrule
    \end{tabular}
    \label{tab:interview_uses_of_gender}
\end{table}
\begin{table}
    \caption{Opinions towards treatment of gender discussed in interviews; frequency represents the number of interviewees that were coded with a particular code.}
    \label{tab:treatment_of_gender_interviews}
    \centering
    \begin{tabular}{@{}lc@{}}
        \toprule
         \textbf{Treatment of Gender Code} & \textbf{Frequency} \\
         \midrule
         Spectrum & 7  \\
         \midrule
         Binary with Other & 3 \\
         \midrule
         Dependent on Audience & 2  \\
         \midrule
         Binary & 1 \\
         \bottomrule \\
    \end{tabular}
\end{table}

Table~\ref{tab:treatment_of_gender_interviews} shows the different ways interviewees stated they would treat gender data. In contrast to Reddit, most interviewees (seven) expressed that gender should be treated as a spectrum---with many discussing that users should be given the opportunity to freely enter any gender, even if other options were given. For example, one participant described the following as how they might collect gender data: ``I guess the easiest way to do it would just to be having [sic] some string input field, like `what gender do you identify as?' And then maybe to write... something like `you don't have to answer this.' Make it an optional field kind of thing, just giving them the option to place whatever they feel in that field like as opposed to giving them a drop down menu or some sort of binary selection.'' While the quoted participant used hedging language like ``I guess'' or ``maybe,'' the participant seemed to have a strong grasp of users' needs. Only one interviewee expressed that gender should be a strict binary, justifying their view by saying that ``some things we need in male and some female.'' Notably, two interviewees were unclear or unsure how they should treat gender data. 
This reflects a general uncertainty which most interviewees had towards our questioning about gender, with interviewees often requiring clarification or time to think before they could discuss what choices users should be given.

This result does not necessarily indicate that participants conceptualize gender in the manner that their responses were coded. The interview questions focused on developer practice---rather than each individual's concept of gender. A participant who indicated that users should be given the option to self-describe their gender---alongside binary choices---may still passively hold to a noninclusive understanding of gender. Additionally, as participants are self-reporting behaviour, their true practice may vary. For example, one participant referred to the ternary gender disclosure field as adding a little ``wokeness,'' in contrast to the binary gender disclosure field. This suggests that, despite reacting positively to the more inclusive gender options, he may have felt that additional gender options were more of a political statement than a way to ensure more users are included.

The expert participant discussed at length the approach to gender taken by their social media community. Rather than requiring users to report their gender identity, the community allowed individuals to report their pronouns. The site also provided the option for users to tag posts with descriptors like ``transwomen''\footnote{The use of a contraction here is due to a tag being a single word. The sense of linguistic othering caused by this contraction is an unfortunate side effect.} so that they could find others who shared their identity on the website. This approach is in line with the best practices discussed in Scheuerman et al.~\cite{Scheuerman_Jiang_Spiel_Brubaker_2021}. It also maximizes users' ability to use the website without unnecessary or uncomfortable disclosure, while also enabling people to find communities which shared their identity.

One aspect of developer views towards gender data which we were unable to evaluate with Reddit was developer thoughts on privacy. While gender is not private for most, gender privacy can be important to marginalized communities and we wanted to see if developers considered such use cases. While not initially part of the interview procedure, the interviewer prompted participants by asking them to think about the privacy of data as a range from the most private thing---like one's tax ID number or bank account password---to completely public things---like a username on a forum. We did not evoke academic models of privacy as we were more interested in developers' immediate perceptions than deeper evaluation.  

Even more so than the previous questions on gender data, developers had to spend time considering their answer. Most (seven) settled on gender information being equivalent to other personal data without special consideration. Four interviewees felt that gender information was entirely public, while two interviewees felt that gender information was extremely private. For example, one of the participants who felt that gender information was extremely private stated: ``I absolutely think gender data is quite private... if you ask me on a scale of one to five, I rate it five, where five is the one I think is the most private data, and I feel people would be more, in terms of the questions that I'd ask, I think gender was the most sensitive one. And that was the reason I had not made it mandatory. People can answer, or they may skip it. It's up to them. But, yes, I completely do believe that gender data is probably one of the most sensitive or private data.'' Finally, one interviewee felt that the privacy of gender information was dependent on audience, specifically mentioning that certain marginalized communities may feel that gender information is extremely private while others might not. Six interviewees also discussed opt-in permissions for gender information, allowing users to choose whether they shared a data point or not. Largely, then, interviewees, while expressing an understanding that personal data requires protection, did not have knowledge of the unique privacy concerns that may surround gender data.

The expert participant discussed their community's general approach to privacy in detail.  They collected almost no private information about users, with profile pictures and usernames being the only potentially directly identifiable information shared by users. Additionally, users had the ability to create ``private'' accounts which were only visible to other users of the platform. This protected users from the public indexing of their posts or profile. Although this was not mentioned by the participant, this likely helped protect users from the data collection present on other websites which can lead to unwilling gendering through advertisers~\cite{Bivens_2017}.

\begin{figure}
    \centering
    \caption{Interviewee reactions to example gender disclosure forms and the ISO/IEC 5218 standard; frequency represents the number of interviewees who reacted in a particular way.}
    \Description[A bar chart with four subcharts, showing that interviewees reacted most positively to the most inclusive form]{A bar chart with four subcharts representing the frequency of each reaction coded in interviews. For the binary form, five interviewees reacted positively, one reacted neutrally, and nine reacted negatively. For the ternary form, four reacted positively, three reacted neutrally, and eight reacted negatively; For the most inclusive form, eight reacted positively, three reacted neutrally, and four reacted negatively. For the ISO/IEC 5218 standard, zero participants reacted positively, four reacted neutrally, and eleven reacted negatively.}
    \includegraphics[width=\textwidth]{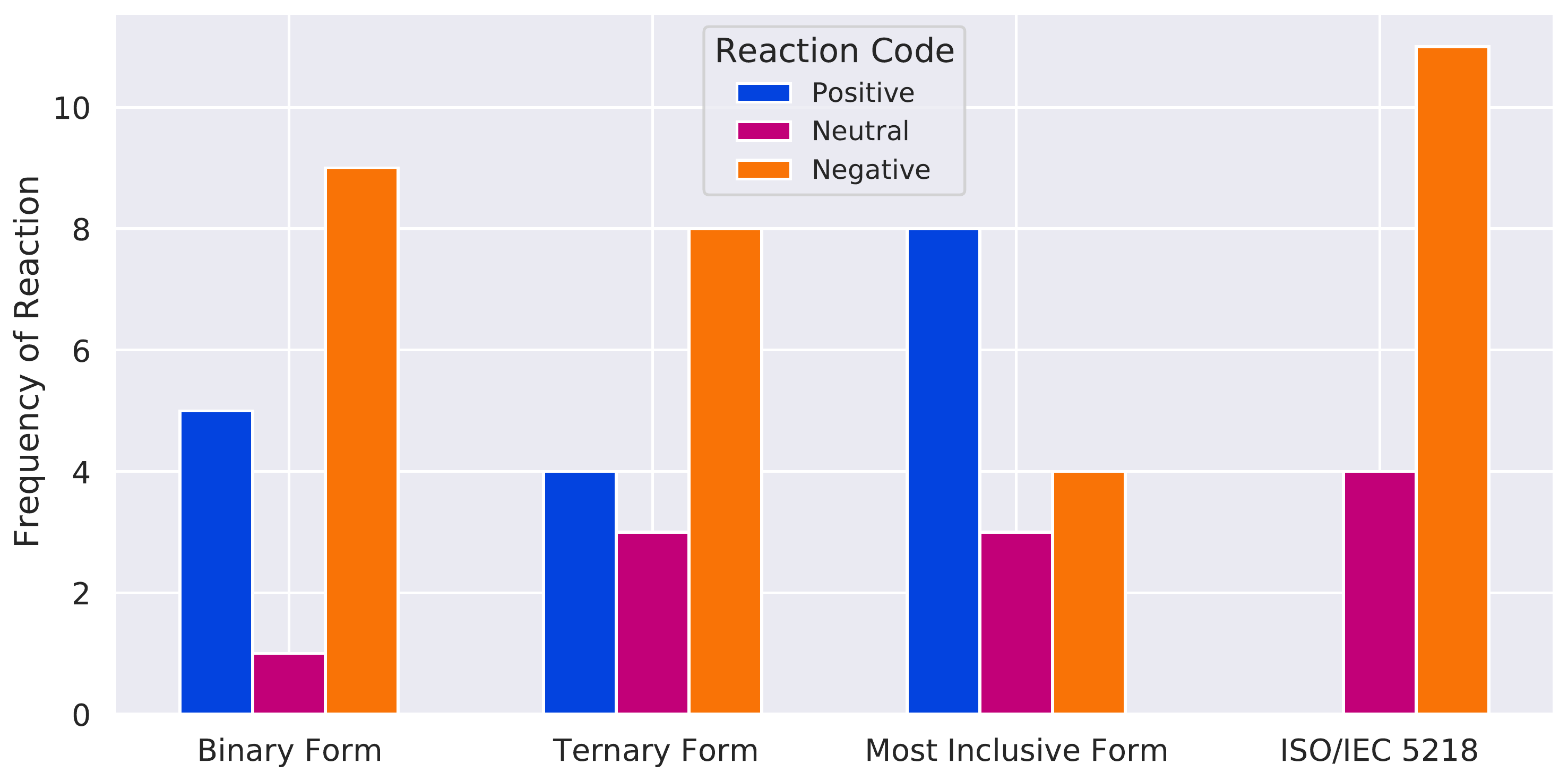}
    \label{fig:overview_reactions}
\end{figure}

The final coded results from our interviews concern developers' reactions to different example gender disclosure forms and the ISO/IEC 5218 standard. An overview of the results can be seen in Figure~\ref{fig:overview_reactions}. Developers reacted most positively to the most inclusive gender selection but did not necessarily have significant critiques for other options. Developers had trouble understanding the second displayed option, which added ``diverse'' to the binary options. This included the developers from Germany who might be expected to be most familiar with the options. The ISO/IEC 5218 standard was uniformly received poorly, with only one of the interviewees having seen the standard previously. Several interviewees correctly pointed out that the standard was for sex rather than gender. Notably, two interviewees---including the developer who had previously worked with marginalized communities---mentioned that while our most inclusive gender option was the best of the three gender disclosure fields, it could be made better by replacing it with a text box. Two interviewees reacted strongly negatively to the most inclusive field, with one stating that identifying with multiple terms is not typical and another stating that people might not want to share if they are non-binary. While this latter concern is valid, it is more of a reason not to collect gender data rather than a reason to use less inclusive forms.

\section{Discussion}
\label{sec:discussion}
Here we discuss the major takeaways of this paper, make recommendations to educators and developers, and suggest future research directions to improve the handling of gender in development.

\subsection{Comparison of Reddit and Interview Study Results}

Returning to our framing research questions, our two studies provide the following results.

\textbf{1. How do developers ask users for gender?} The plurality of our interviewees showed a preference for treating gender as a spectrum. Additionally, the most inclusive gender disclosure forms presented in interviews received the most positive reactions. In contrast, in our Reddit dataset, the majority of analyzed code snippets relied on a binary conception of gender. Based on the totality of our evidence, as well as knowledge of prior research, we find that programmer practice\footnote{And not an insignificant amount of educational material.} still relies on binary disclosure forms. Our interviews may indicate increasing acceptance of more inclusive disclosure forms, but this result may also be an artifact of the law of small numbers, bias in recruitment, or other biases in study techniques (see the ``Limitations'' portion of Subsection~\ref{ssec:interview_method}).

\textbf{2. How do developers use user gender data?} The two studies speak most to the initial purposes of gender data. Audience demographic analysis was both the most common use case found in our Reddit dataset and the most common use mentioned by interviewees. Artificial intelligence and machine learning uses of gender were the next most common use cases found on Reddit, followed by health or fitness applications. Interviewees more often mentioned dating applications, with only two interviewees mentioning artificial intelligence and machine learning as uses for gender data. These results suggest that gender data is used in a variety of situations, including multiple ways which may not be essential to the primary function of an application. Moreover, any utility gained by collecting gender to understand audience demographics, for example, is eroded when an overly narrow, inaccurate disclosure form is used.

\textbf{3. What advice do developers receive about using gender data?} This research question was primarily addressed using our Reddit data. From the (admittedly small) amount of discussion of gender disclosure and data use that we observed, users who requested advice received inclusive recommendations. Beyond the specific context of asking for advice, inclusive notions of gender seem to be viewed with hostility on Reddit. Overall, then, the treatment of gender data seems to be a topic that is largely not discussed by programmers in the communities we observed. This conclusion is consistent with the apparent uncertainty of many of our interviewees in answering some of our questions regarding gender.

Overall, then, while there are positive signs of more inclusive practices, mainly among our interviews, the public conversation and practice found on Reddit suggest that more work is necessary to improve how developers approach gender in software. The binary conception of gender seems to be dominant, at least on Reddit. This finding aligns with the experiences of non-binary individuals with gender representation in software and prior research on the topic~\cite{10.1145/3461778.3462033, Scheuerman_Jiang_Spiel_Brubaker_2021, 10.1145/3290607.3310425}.
The common usage of gender data for audience demographics---seen in practice on Reddit and mentioned commonly by interviewees---suggests that developers view gender information as an important data point for categorizing users. This practice is almost certainly influenced by the demands of online monetization, with gender-related code on Reddit including explicit references to advertising services. Prior literature~\cite{Bivens_Haimson_2016, Bivens_2017} has identified how advertising considerations have influenced the design of gender categorization in social media services. Demographic classification along the lines of gender collapses the nuance inherent in one's identity, especially when only two categories are employed.

\subsection{Towards Better Gender Programming Practices}

How might developers handle gender data better? The answer in many cases may be to avoid encoding gender data altogether. Particularly in contexts where gender data is non-essential to the functioning of a system, this approach would help to avoid the pitfalls associated with gender categorization. For example---while behavioral advertising has its associated harms~\cite{boerman2017online}---it makes far more sense to target ads based on purchasing habits than to target ads based on gender data. The former avoids the assumptions and stereotypes intrinsic to the latter. For social media platforms, it makes the most sense to only collect pronouns~\cite{Scheuerman_Jiang_Spiel_Brubaker_2021}. As described by our expert participant, this need not prevent the formation of community based on gender identity via user-generated content. In the domain of computer vision, Scheuerman et al. recommend that image labeling systems ``embrace gender ambiguity'' and label gender-neutral features like ``beard, makeup, dress'' rather than attempting to force all humans into a false gender binary~\cite{scheuerman2019computers}.

Importantly, the commitment to removing gender must be more than skin-deep. Removing users' option to disclose their gender identity while still categorizing them into a binary on the back end serves only to obfuscate how gender is ``baked'' into a  system~\cite{Bivens_Haimson_2016}. Additionally, gender prediction is likely to exclude non-binary individuals and generally risks misgendering individuals~\cite{hamidi2018gender,scheuerman2019computers, scheuerman2020we}.

It is not necessary to exclude gender from computing altogether. While there does not exist a single, all-encompassing approach to gender in computing, prior work on gender inclusivity in HCI research provides some guidance. As described in Section~\ref{sec:background}, gender is not a single concept but rather many overlapping ideas which vary with context and time~\cite{10.1145/3449113}. Even a well-designed gender disclosure form\footnote{See guidelines G-3 through G-6 of Scheuerman et al.~\cite{scheuerman2020hci} or Spiel et al.~\cite{10.1145/3338283}} is necessarily limited to its context. Developers should consider what definition of gender they are using and what measure best corresponds to this definition~\cite{10.1145/3449113, scheuerman2020hci}. Users should be given enough information to understand what is being asked of them and how it will be used~\cite{10.1145/3449113, scheuerman2020hci}.  Developers should not take a single measure out of context and extrapolate about other aspects of gender. For example, the pronouns one uses should not be used to make assumptions about one's gender identity as done on Facebook~\cite{Bivens_Haimson_2016, scheuerman2020hci}. Similarly, if inclusive options are given, they should not be purely aesthetic. That is to say, it is not appropriate to collect non-binary gender data but then simply exclude those people from analysis or normal treatment for not fitting within the binary paradigm. Returning to the example of online advertising and social media, Facebook's choice to allow free-form gender disclosure is good, but it is undermined by the practice of essentially ignoring non-binary individuals in ad-targeting~\cite{Bivens_Haimson_2016}.

This approach to gender is more complicated than declaring \verb|bool gender| and calling it a day. We freely confess that an inclusive approach will require more effort than any reductionist approach. Further, adoption will likely require those with power to advocate for those who are ignored by the binary gender paradigm. These hurdles do not excuse inaction, however, as all users deserve systems that do not deny their existence. To quote Spiel et al. ``What we cannot do is simplify [gender] or say that non-binary inclusion is just an additional checkbox, or even that a one size fits all solution exists for a population whose very existence denies the idea of simple fixes or classifications. There is no easy, single answer here, but that the work is hard is not a
reason to avoid''~\cite{10.1145/3290607.3310425}.

In addition to our recommendations for developers, we would also suggest that change is necessary in the area of software engineering education. While individuals may expand their understanding of inclusive practices through private study, course designers should consider explicitly discussing inclusive practices for gender data as part of their curricula.\footnote{One approach would be to integrate such discussions into a course focused specifically on diversity, equity, and inclusion in computer science. For example, a group of Ph.D. candidates at Carnegie Mellon University have successfully developed and implemented such a course~\cite{Aupperlee2022}.} We understand that course time is limited, but by doing so, educators could help ensure that their students enter the workforce with a nuanced understanding of gender. At the very least, educators should avoid using exclusively binary gender forms in coursework or examples. On Reddit, we observed several examples of code from coursework employing exclusively binary gender disclosure forms. Such practices normalize cisgenderism and perpetuate the bad designs that motivated our research.

Finally, it is worth considering the role of large technology companies in determining how gender data is handled. In addition to being influential through their sheer prominence, the high degree of interdependence in modern software means that decisions made about gender data on a platform may have an inordinate impact on other systems. For example, Bivens et al.~\cite{Bivens_Haimson_2016} identified how the central role of social media services in online advertising may offer them the ability to shape how advertisers view gender. In line with this conclusion, we identified five cases in our Reddit dataset wherein code-snippets used gender data pulled from Facebook or Instagram. While this finding in no way abrogates the responsibility of smaller companies and individual developers to handle gender appropriately, it does demonstrate how pushing more inclusive practices at the largest software industry players may result in a cascade of change. Systemic change is, of course, easier discussed than done, but we must fight for it if we are to make computing accessible for all.

\subsection{Future Work}

Before concluding, we would like to outline some potential directions for future research on gender and software development. While we feel our findings stand on their own, we encourage work that examines the same or similar questions on a larger scale---for example, through a widely distributed survey. Another option would be to systematically analyze practice through publicly available repositories of code. Looking at the open source community (e.g. using World of Code~\cite{ma2019world}) may also provide the ability to analyze how such decisions are made through the associated discussions~\cite{8147104}. Commercially available software might be analyzed through static analysis.  Such research would offer a wider perspective on how developers handle gender data.  These methods may also be used to explore the secondary usage of gender data.  While both our interview and Reddit study suggest that audience demographics are the primary purpose for gender data in software, this almost certainly is not the end of the story. Without further study, we cannot directly comment on how data may be reinterpreted into a binary paradigm despite being initially collected in more inclusive ways.

Lastly, it is important to investigate the best methods for increasing inclusive gender practices among developers. We have made some general suggestions in the preceding subsection about software engineering education, but deliberate research is necessary. One direction would be to create educational interventions which teach about issues of gender inclusivity as has been done with other issues of diversity, equity, and inclusion~\cite{10.1145/3408877.3432557}. Another fruitful line of research may be to develop tools that assist developers by finding gender inclusivity issues---just as technology is used to identify other accessibility issues~\cite{10.1145/3218585.3218673}. Importantly, this research must also be paired with advocacy and more explicit discussion of concepts of gender in computing.\footnote{Crucially, this advocacy must not be left to non-binary people or other marginalized individuals alone. No one should have to continually justify their existence.} As long as the binary paradigm remains unchallenged, development practices will not change.
\section{Conclusion}
\label{sec:conclusion}

In this paper, we have presented a mixed-methods study investigating how developers request gender disclosure and use gender data. Through our analysis of both interviews and Reddit posts, we found that developers rarely discuss or receive advice on inclusive practices with regards to gender information. We also found that much of the code posted to Reddit uses a strict binary conception of gender. These failures contribute to the systematic, technological erasure of non-binary and other gender non-conforming individuals. Ultimately, we hope that our findings serve as a platform not only for increased scrutiny of how developers treat gender but also for deeper discussion on concepts of gender. Such work is essential in order to improve the lives of transgender, non-binary, and other gender non-conforming people. 
\begin{acks}
The authors would like to thank Marcel Fourné, Dr. Lucy Simko, and David Mayes for their support of this work. The authors would also like to thank MPI-SP for supporting this research in many ways, including infrastructure and a remote internship. 
\end{acks}

\bibliographystyle{ACM-Reference-Format}
\bibliography{bibliography}
\received{January 2022}
\received[revised]{July 2022}
\received[accepted]{November 2022}

\clearpage
\appendix
\section{Screener Survey }
\label{app:screener_survey}

\textbf{Introduction Text}: 

We will now ask some questions to confirm that you are eligible for the study. We also will ask for demographic information. We use this information to be able to describe differences between groups and to increase the visibility of less represented groups. Answers, with the exception of those questions directly related to study eligibility, will not be used to determine if you can continue in this study or not. Please answer as honestly as possible.

\textbf{Questions}:

\begin{enumerate}
    \item What is the highest level of school you have completed or the highest degree you have received? 
        \begin{enumerate}
            \item Less than high school degree (Primary school or some secondary school)
            \item High school graduate (high school diploma, equivalent including GED or secondary school)
            \item Some time at college/university but no degree (some tertiary education)  
            \item Technical school or apprenticeship  
            \item 2-year degree at college/university
            \item 3-4 year degree at college/university (bachelor's degree or equivalent)
            \item Some graduate work without degree (graduate certificate or non-degree courses)
            \item Master's degree
            \item Doctoral degree
            \item Professional degree (JD, MD)
            \item Other (Please specify): (\textit{Text entry})
        \end{enumerate}
    \item If you indicated that you received a degree, what was the field of study? \textit{Text entry}
    \item Which statement best describes your current employment status?
        \begin{enumerate}
            \item Employed full-time
            \item Employed part-time
            \item Independent contractor, freelancer, or self-employed
            \item Not working (temporary layoff from a job)
            \item Not working (looking for work)
            \item Not working (retired)
            \item Not working (disabled)
            \item Other (Please specify): (\textit{Text entry})
            \item Prefer not to answer
        \end{enumerate}
    \item How many years of experience do you have in the field of software development/programming/coding? (\textit{Text entry})
    \item How did you learn programming? Check all that apply
        \begin{enumerate}
            \item University Courses
            \item Textbooks/Academic Books
            \item Research Papers
            \item Tutorial Books
            \item Online courses (e.g., Coursera)
            \item Informal Community Discussion (Discord, Slack)
            \item Forum-based platforms (e.g. Reddit)
            \item QA platforms (e.g. Kaggle, Stack Overflow)
            \item Company-provided training
            \item Company-Sponsored Conferences
            \item Other (Please specify): (\textit{Text entry})
        \end{enumerate}
    \item What is your age? (If you prefer not to answer, enter a number greater than 150) (\textit{Text entry})
    \item Which of the following best describe you? Please select all that apply.
        \begin{enumerate}
            \item Black or of African descent
            \item East Asian
            \item Hispanic or Latino/a/x
            \item Indigenous (such as Native American, Pacific Islander, or Indigenous Australian)
            \item Middle Eastern
            \item South Asian
            \item Southeast Asian
            \item White or of European descent
            \item Self Description (\textit{Text entry})
            \item Prefer not to answer
        \end{enumerate}
     \item Which term best describes your gender identity? If you identify with multiple labels, please select "self identify" and list all that apply
        \begin{itemize}
            \item Man
            \item Woman
            \item Non-Binary
            \item Self identify (\textit{Text entry})
            \item Prefer Not to Answer
        \end{itemize}
    \item In which country do you currently reside? (\textit{Dropdown box})
\end{enumerate}

\section{Interview Procedure}
\label{app:interviewprocedure}

\subsection{Pre-interview Procedure}
\label{app:sub:preinterview}
\begin{enumerate}
\item \textbf{Introduction}: Introduce yourself e.g.: \textit{``Good afternoon! Thank you for coming and participating in our interview study. I am...''}
\item \textbf{Overview}: Describe study and procedure, e.g.: \textit{``We’re going to be asking you some questions today about your experiences working on programming projects with specific regard to gender.''}
\item \textbf{Consent}: Reference the consent form, remind them of confidentiality and opt-out e.g.: \textit{``When taking our survey you signed a consent form, I just wanted to emphasize  that what you say to us during the interview will be kept confidential, and you can stop participating at any time, just let us know.''}
\item \textbf{Recording}: State that you will now turn on the recording, then do so, tell them it is now running
\item \textbf{Interview}: Ask the interview questions, remember not to prime, to listen carefully and follow up where appropriate.
\end{enumerate}
\subsection{Interview Questions}
\label{app:sub:interview_questions}
\begin{itemize}
    \item How would you describe the type of development work you do?
    \item Have you worked with personal data before (e.g. users’ phone numbers)? Explain.
    \item Have you worked with gender data before? Explain.
    \item How have (would) you store gender data? How have (would) you ask about gender?
    \item Do you think gender data is private? To what degree?
    \item Are there any particular concerns you would have in programming or designing the UX for a system which used gender data?
    \item How would you acquire gender information from users for a project? When or how would you ask?
    \item \textit{Show example inclusive and non-inclusive gender options (see Appendix~\ref{sub:app:example_gender}). For each, ask} What do you think of this option? Is it appropriate or useful?
    \item Do you have have any other thoughts you would like to share?
\end{itemize}

\subsection{Example Gender Options}
\label{sub:app:example_gender}
\begin{enumerate}
    \item \textbf{Example Question:} What is your gender?
        \begin{itemize}
            \item Male
            \item Female
            \item Prefer Not to Answer
        \end{itemize}
    \item \textbf{Example Question:} What is your gender?
        \begin{itemize}
            \item Male
            \item Female
            \item Diverse
            \item Prefer Not to Answer
        \end{itemize}
    \item \textbf{Example Question:} Which term best describes your gender identity? If you identify with multiple labels, please select "self identify" and list all that apply
        \begin{itemize}
            \item Man
            \item Woman
            \item Non-Binary
            \item Self identify (Text box)
            \item Prefer Not to Answer
        \end{itemize}
    \item \textbf{ISO/IEC 5218 Information technology:} Codes for the representation of human sexes
    The four codes specified in ISO/IEC 5218 are:
        \begin{itemize}
            \item 0 = Not known;
            \item 1 = Male;
            \item 2 = Female;
            \item 9 = Not applicable.
        \end{itemize}
    The standard specifies that its use may be referred to by the designator "SEX".
\end{enumerate}

\section{Interview Codebook}
\label{app:codebook}

The codebook for interviews is below, divided across two pages. The far left column indicates the overall category of the codes in each subsection. The middle column indicates the code being defined. The far right column lists a description of the code. There are 8 overall categories of codes.

\begin{table}[H]
\begin{tabular}{p{0.18\linewidth}p{0.17\linewidth}p{0.6\linewidth}}
\toprule
\textbf{Code Category} & \textbf{Code} & \textbf{Definition} \\ \midrule
Developer Type & Backend & Developer indicated that they focus on non-user facing aspects of an application \\ \midrule
 & Frontend & Developer indicated that they focus on user-facing aspects of an application or web page \\ \midrule
 & Fullstack & Developer indicated that they do both frontend and backend development \\ \bottomrule
Experience with Gender Data & Yes & Developer indicated that they had previously implemented applications that used gender data \\ \midrule
 & No & Developer indicated that they had not previously implemented applications that used gender data \\ \bottomrule
Use of Gender & Audience or Client Demographics & Gender used to get insight about the makeup of their clients or application audience \\ \midrule
 & Diversity Protection & Gender used to ensure users from diverse backgrounds receive equitable treatment \\ \midrule
 & Dating & Gender used to coordinate preferences on dating applications or websites \\ \midrule
 & AI/ML & Gender used for artificial intelligence or machine learning \\ \midrule
 & Healthcare & Gender used for healthcare purposes, like intake or prescriptions \\ \midrule
 & 3rd Party Telemetry/Advertising & Gender used to enable advertising functionality in applications or websites through a third party service like Google. \\ \midrule
 & Legal or Financial Purposes & Gender used to enable legal or financial services, like banking or insurance. \\ \midrule
 & No Specific Purpose & Gender collected without a specific purpose for its use \\ \bottomrule
Experience with Personal Data & Yes & Developer indicated that they had previously implemented applications that used a user's personal data \\ \midrule
 & No & Developer indicated that they had not previously implemented applications that used a user's personal data \\ \bottomrule
 \end{tabular}
 \end{table}
\begin{table}[H]
\centering
\begin{tabular}{p{0.18\linewidth}p{0.17\linewidth}p{0.6\linewidth}}
\toprule
\textbf{Code Category} & \textbf{Code} & \textbf{Definition} \\ \midrule
Treatment of Gender & Dependent on Audience & Developer indicated they would chose how to treat gender based on the intended audience \\ \midrule
 & Binary & Developer indicated they would use ``male'' and ``female'' gender options \\ \midrule
 & Binary with Third Option & Developer indicated they would use ``male'' and ``female'' gender options with a third option like ``other'' \\ \midrule
 & Spectrum & Developer indicated they would treat gender as a spectrum in some way, such as by including a text input for gender. \\ \bottomrule
Gender Privacy & Equal to Other Personal Data & Developer indicated gender is as private as other user personal data \\ \midrule
 & Opt-in permissions & Developer indicated they would allow users to opt-in to sharing gender \\ \midrule
 & Public & Developer indicated that gender is public data \\ \midrule
 & Dependent on Audience & Developer indicated that how private gender is depends on the intended audience. \\ \midrule
 & Most Private & Developer indicated that gender is among the most private data \\ \bottomrule
Reaction to Gender Selection & Positive & Developer reacted positively to a particular example gender selection. \\ \midrule
 & Neutral & Developer reacted neutrally to a particular example gender selection; they may have offered some light critiques and/or a small way to change the form. \\ \midrule
 & Negative & Developer reacted negatively to a particular example of gender selection; they offered a serious critique and/or multiple ways to change the form. \\ \bottomrule
\end{tabular}
\end{table}

\typeout{} 
\end{document}